\documentclass{aa}
\usepackage{graphics,epsfig}

\def\simless{\mathbin{\lower 3pt\hbox
     {$\rlap{\raise 5pt\hbox{$\char'074$}}\mathchar"7218$}}}   
\def\simmore{\mathbin{\lower 3pt\hbox
     {$\rlap{\raise 5pt\hbox{$\char'076$}}\mathchar"7218$}}}   

\begin{document}

\title{Spectra and time variability of galactic black-hole X-ray sources
in the low/hard state }

\subtitle{}

\author
{Giannios D.\inst{1,2}, Kylafis N. D.\inst{1,2}, Psaltis D.\inst{3}}

\institute{
University of Crete, Physics Department, P.O. Box 2208, 710 03,
Heraklion, Crete, Greece
\and Foundation for Research and Technology-Hellas, 711 10, Heraklion,
Crete, Greece
\and University of Arizona, 1118 E. 4th St., Tucson, AZ 85721, USA
}

\authorrunning{Giannios et al.}
\titlerunning{Spectra and variability of black-hole sources
in the low state}

\offprints{dgiannios@physics.uoc.gr}

\date{Received: \\
Accepted: \\}

\abstract{

We propose a jet model for the low/hard state of galactic black-hole X-ray sources which
explains a) the X-ray spectra, b) the time-lag spectra, c) the increase of the
amplitude of variability (QPO and high frequency) with increasing photon energy, and 
d) the narrowing of the autocorrelation function with increasing photon energy.  The 
model (in its simplest form) assumes that i) there is a uniform magnetic field along the
axis of the jet, ii) the electron density in the jet is inversely proportional to 
distance and iii) the jet is ``hotter'' near its center than at its periphery.  We have 
performed Monte Carlo simulations of Compton upscattering of soft photons from the 
accretion disk and have found power-law high-energy spectra with photon-number index
in the range 1.5 - 2, power-law time lags versus Fourier frequency  with index
$\sim 0.8$, and an increase of the rms amplitude of variability and
a narrowing of the autocorrelation function with photon energy as they have been
observed in Cygnus X-1. Similar energy spectra and time variability have been 
observed in neutron-star systems in the island state.  It is therefore quite likely that a
similar model holds for these sources as well.

}

\maketitle

\section{Introduction} \label{intro}
 
Galactic black-hole binaries have been extensively observed for more than
thirty years. Their spectral and timing behavior exhibits a great 
variety of properties, yet in most cases they can be classified to be
in one of few different states. The X-ray spectra of black-hole candidates
have two distinct components: a multi-color black-body component (Mitsuda et
al. 1984, but see Merloni et al. 2000), widely accepted to come from the 
accretion disk (Shakura \& Sunyaev 1973), and a non-thermal hard X-ray tail 
that is usually modeled to come from unsaturated 
inverse Compton scattering of soft photons with energetic electrons 
(see e.g. Sunyaev \& Titarchuk 1980; Poutanen 2001).   
The relative strength of the thermal component in comparison with the hard
tail, the slope of the tail, and the variability are used to
identify the spectral state of the source at a given
epoch. When the luminosity of the source is low (less than a few percent of the
Eddington luminosity) and the thermal component is weak or absent, with the 
non-thermal tail dominant and the rms variability of the source high (20\% - 
50\%), then the source is in the low/hard state (see e.g. van der Klis 1995). 

The location of the energetic electrons that Comptonize a soft photon
input (and result in the power-law emission) is model dependent.
One possibility is that hot, optically thin plasma exists in a corona 
(Galeev et al. 1979; Haardt \& Maraschi 1993; Haardt et al. 1994; 
Stern et al. 1995; Poutanen \& Fabian 1999) that 
lies above and below the thin disk. A second possibility is a two-temperature 
flow (Shapiro et al. 1976; Ichimaru 1977; Rees et al. 1982;
Narayan \& Yi 1994; Abramowicz et al. 1995; Esin et al. 1998).

There is growing evidence that all X-ray binaries harboring a black hole
display radio emission when they are in the low/hard X-ray state (Hjellming 
\& Han 1995; Fender et al. 1997; Mirabel et al. 1998; Fender 2001; 
Stirling et al. 2001; Gallo et al. 2003). Inverse Compton scattering by 
relativistic electrons in a jet has been proposed as a mechanism for the 
production of X-rays and $\gamma$-rays in Active Galactic Nuclei 
(Begelman \& Sikora 1987; Bednarek et al. 1996; Harris \& Krawczynski 2002) and 
X-ray binaries (Band \& Grindlay 1986; Levinson \& Blandford 1996; Georganopoulos et
al. 2002; Romero et al. 2002). The contribution of the synchrotron
emission from the jet to the hard X-rays might also be significant
(Markoff et al. 2001; Corbel \& Fender 2002).

Spectral information alone, however, can not reveal the accretion geometry.
This is so because rather different models can result in very
similar spectra. So as to make further progress in our understanding of
how accretion works, studies of the timing properties of these sources are also 
essential. 

The fact that phase lags between two different X-ray
bands are approximately independent of Fourier frequency (Miyamoto et al.
1988) is impossible to explain with models where the Comptonization is taking 
place in a hot, uniform, optically thin plasma. 
This led to the idea of a large hot Comptonizing medium with density inversely
proportional to radius (Hua et al. 1997; Kazanas et al. 1997; Kazanas \& Hua 1999;
Hua et al. 1999). Still this model failed to explain the fact that the 
autocorrelation function of the light curve of Cygnus X-1 in the low/hard state
becomes narrower as the photon energy increases (Maccarone et al. 2000).   

Recently, Reig et al. (2003) proposed that inverse Compton scattering of
soft disk photons by energetic electrons in a jet can explain both
the X-ray emission of black-hole candidates in the low/hard state and the 
decreasing time lags with Fourier frequency. 
The density of the electrons in this
(quite simple) jet model is assumed to drop inversely proportional of the
distance $z$ from the black hole. 
In such a case, and for an effective optical depth to electron scattering
of order unity along the axis of the jet, there is equal probability for
scattering of the photons per decade of distance $z$.  The smallest 
period of variability that can pass through the scattering medium is
of the order of the timelag caused by scattering, which is of the order
of $z/c$.  Faster variability
gets smoothed out.  This means that the maximum frequency of variability
that can pass through the jet is inversely proportional to the timelag.
Thus, timelags roughly inversely proportional to Fourier frequency can be 
produced.  Applying this model to the case of Cygnus X-1 it was possible 
to reproduce both the energy and the time-lag spectra. 

In this work we make one simple modification to the model described by
Reig et al. (2003).  Namely, we assume that the electrons are more 
energetic at small polar distances $\rho$ (in the core of the jet) than
at larger $\rho$ (jet's periphery). With this modification we 
are able to explain two additional observational facts exhibited by Cygnus X-1. 
One is the narrowing of the autocorrelation function with increasing photon energy
(Maccarone et al. 2000) and the other is the hardening of the high-frequency power 
spectrum with increasing photon energy (Nowak et al. 1999). 
Preliminary results of this work appeared in Kylafis et al. (2004).
A similar approach was followed by Lehr et al. (2000) to explain the energy 
dependence of the fractional amplitude of the high-frequency QPOs in GRS 1915+105.              

\section{The model}
\subsection{Characteristics of the jet} 

Our jet model is quite similar to the one described in Reig et al. (2003). 
More specifically, we assume that the jet is accelerated
close to its launching area and maintains a constant (mildly 
relativistic) velocity thereafter.  Furthermore, and for computational 
simplicity, we assume that a uniform magnetic field exists along the 
axis of the jet (taken to be the $z$ axis). The electrons' velocity has 
two components one parallel ($v_{\parallel}$) and one perpendicular 
($v_{\perp}$) with respect to the magnetic field. 

The new ingredient of the model is that the perpendicular component of the
velocity is not constant, but it drops linearly with polar distance 
$\rho$, with its value in the outer edge of the jet being a fraction 
(typically half) of that along the $z$ axis. The result is a jet whose 
core is ``hotter'' than its periphery. This,
rather reasonable, assumption will have important implications on the 
shapes of the autocorrelations and the power-density spectra of the 
light curves in different energy bands.

The electron-density profile in the jet is taken to be of the form

$$
n_{\rm e}(z) = n_0 (z_0/z)^p,
\eqno(1)
$$         
where $z$ is the vertical distance from the black hole and $p$ is a parameter
close to 1 (see Sect. 3 below).  The parameters
$n_0$ and $z_0$ are the electron density and the height at the base of the
jet respectively. If $H$ is the extent of the jet along the $z$ axis, 
then for $p=1$ the Thomson optical depth along the axis of the jet is

$$
{\tau}_{\parallel}=n_0{\sigma}_T z_0 \ln (H/z_0).
\eqno(2)
$$

Let $\pi R^2(z)$ be the cross-sectional area of the jet as a function
of height $z$. Then
from the continuity equation $\dot M=\pi R^2(z)m_pn_e(z)v_{\parallel}$,
we obtain the radial extent $R$ of the jet as a function of height $z$

$$
R(z)=(R^2_0z/z_0)^{1/2}.
\eqno(3)    
$$
Here $R_0=(\dot M/\pi m_pn_0v_{\parallel})^{1/2}$ is the radius of the jet
at its base, $\dot M$ is the mass-ejection rate and $m_p$ is the proton
mass.

\subsection{The Monte Carlo simulations} \label{code}

For our Monte Carlo simulations we follow Cashwell \& Everett (1959) and
Pozdnyakov et al. (1983).  A similar to ours, albeit independent, code was used 
in the work of Reig et al. (2001, 2003).  The two codes are in excellent 
agreement between them.  The procedure we follow is described briefly below.

Photons from the accretion disk, in the form of 
a blackbody distribution of characteristic temperature
$T_{\rm bb}$, are injected at the base of the jet with an upward
isotropic distribution.  It is not clear how the blackbody temperature
of the disk varies with polar distance $\rho$, since we do not have a pure 
accretion disk, but we also have ejection of matter from the disk in the form of 
a jet.  Thus, for simplicity, we consider a constant $T_{\rm bb}$ 
and constant luminosity per unit polar distance $\rho$.  These assumptions are 
not crucial for our conclusions.  
We also consider disk photons that are emitted at poloidal distances $\rho < R_0$. 
This is justified by the fact that, for the typical values of $R_0$ 
considered here ($\sim 10^8$ cm, see Sect. 3), the disk
flux outside this radius is small and furthermore, only 
a small fraction of it will interact with the jet. 
 
Each photon is given a weight equal to unity when it leaves the accretion 
disk and its time of flight is set equal to zero.  The optical depth
$\tau (\hat n) $ along the photon's direction $\hat n$ is computed and a 
fraction $e^{- \tau (\hat n)}$ of the photon's weight escapes and is
recorded.  The rest of the weight of the photon gets scattered in the jet.
If the effective optical depth in the jet is significant (i.e., 
$\simmore 1$), then a progressively smaller and smaller weight of the photon
experiences more and more scatterings.  When the remaining weight in a
photon becomes less than a small number (typically $10^{-8}$), we start 
with a new photon.  

The time of flight of a random walking photon (or
more accurately of its remaining weight) gets updated at every scattering
by adding the last distance traveled divided by the speed of light.  For the
escaping weight along a travel direction we add an extra time of flight
{\it outside} the jet in order to bring in step all the photons (or fractions
of them) that escape in a given direction from different points of the 
``surface'' of the jet.  The more a fractional photon stays in the jet,
the more energy it gains on average from the circular motion
(i.e., $\vec{v_{\perp}}$) of the electrons.  Such Comptonization
can occur everywhere in the jet.  Yet, a photon that random walks high up
in the jet has a larger time of flight than a photon that random walks
near the bottom of the jet.

If the defining  parameters of a photon (position,  direction, energy,
weight, and time of flight) 
at each stage of its flight are computed, then we can determine
the spectrum of the radiation  emerging from the  scattering medium and
the time delay of each escaping fractional photon. The optical depth to 
electron scattering, the energy shift, and the new direction of the 
photons after scattering are computed using the corresponding 
relativistic expressions.

\subsection{The high-frequency power spectrum}
\label{power}

The power spectra of Cygnus X-1 in the low/hard state can be 
well fitted by a doubly broken power low 
(Churazov et al. [2001]; see, however, Nowak [2000], Belloni et al. [2002] and 
Pottschmidt et al. [2003] for a different approach and Revnivtsev et al. 
[2000] for evidence of a break above $\sim 40$ Hz). 
Other black-hole candidates show similar power spectra when they
are in the low/hard state (Miyamoto et al. 1992).
The first break is typically observed at frequencies 
of the order of 0.1 Hz, where the slope changes from zero to 
$\sim -1$ (when the power is expressed in units of $rms^2$/Hz) and the 
second break at about 1 Hz with the high Fourier-frequency slope being $\sim -1.5$.

Nowak et al. (1999) showed, for the case of Cygnus X-1, that while the slopes 
of the first two power laws (slopes $\sim 0$ and $\sim -1$) do not depend on 
photon energy, the high-frequency power law exhibits a characteristic hardening  
with increasing photon energy. In this work we show that this hardening can be 
explained by a jet that has a core ``hotter'' than its periphery.    
Note, however, that other black-hole candidates show an energy independent
power-density spectral shape (Lin et al. 1999).  In the context of our model,
this means that the jets in these sources have $v_{\perp}(\rho) \approx $ constant.

In order to compute the power spectrum of the light curves in different energy
bands, further assumptions concerning the source of the X-ray variability
have to be made. We assume that the variability originates in the accretion disk,
which is the source of the soft-photon input.  This variability is modified 
as the disk photons travel through the jet and create the power-law energy
spectrum.

We think that it is reasonable to assume that
the highest frequency of variability, produced in a ring of disk material at
radius $\rho$, is of the order of the Keplerian frequency at this radius,
namely 
$$
\nu_{\rm K}(\rho)=1/(2\pi)(GM/\rho^3)^{1/2},
\eqno(4)
$$ 
where $M$ is the mass of the black hole. This assumption is justified by the 
small number of characteristic frequencies that describe the power spectra 
(Belloni et al. 2002). 
This then implies that a Fourier frequency $\nu$ is associated 
with variability in the part of the accretion disk with radius  
$$
\rho < (2\pi\nu)^{-2/3}(GM)^{1/3}.
\eqno(5)
$$

On the other hand, we have assumed that the core of the jet is ``hotter'', 
resulting in a harder energy spectrum for input soft photons
that come from the region close to the black hole.  This is because these 
photons sample the core of the jet more than its periphery.
Thus, higher Fourier frequencies are naturally associated with harder energy
spectra, making the power spectra flatter in the harder energy bands. 
This is exactly what is observed in the case of Cygnus X-1 (Nowak et al. 
1999) and it will be quantitatively presented in Sect. \ref{powersp}.
Furthermore, in the context of this model, the break of the power
spectrum that occurs at a frequency of a few Hz is naturally attributed to
the Keplerian frequency at the outer edge $\rho=R_0$ of the base of the jet.

\section{Parameter values}

Our jet model has a number of free parameters.  The values of these parameters
are chosen so that they are reasonable and they produce results similar to the
ones observed.  

The parameters of our model are the following: The exponent $p$ (Eq. [1]), 
the temperature $T_{bb}$ of the soft photon input, 
the extent $H$ of the jet,
the radius $R_0$ and the height $z_0$ of the base of the jet, 
the Thomson optical depth ${\tau}_{\parallel}$ along the axis of 
the jet, the parallel component of the velocity 
$v_{\parallel}$ of the electrons, the perpendicular velocity component 
$v_{\perp,0}$  at $\rho = 0$,  and its value at the outer 
edge ($\rho = R$) of the jet 
$v_{\perp,{\rm out}}$. We assume that $v_{\perp}$ drops linearly with radius 
$\rho$,
but any other smooth drop gives qualitatively similar results.  
We remark here that, due to the mildly relativistic flow in the jet,
the emerging X-ray spectra from the jet depend on the angle between 
the line of sight and the magnetic field direction.

The values of the parameters that reproduce quite well the observations of
Cygnus X-1 are here called {\it reference values} and they are:
$p=1$, $kT_{\rm bb} = 0.2$ keV, $H=10^5 r_g$, 
$R_0=100 r_g$, $z_0=5r_g$, $\tau_{\parallel}=2.5$, $v_{\parallel}=0.8 c$, 
and $v_{\perp,0}=0.55 c=2 v_{\perp,{\rm out}}$, 
where $r_g = GM/c^2 = 1.5 \times 10^6$ cm stands for 
the gravitational radius of a 10 solar-mass black hole.    

In order to have good statistics in our Monte Carlo results, we combined all
the escaping photons with directional cosine with respect to the axis of the jet
in the range $0.2< \cos \theta < 0.6$. 
Strictly speaking, this is inconsistent with our assumption of a uniform magnetic
field.  If the magnetic field were uniform, then only one value of $\cos \theta$ 
would be appropriate for each source.  As we will see (Sect. 4), a range of
directional cosines is absolutely necessary in order to reproduce the observed
phenomena.  Therefore, a non-uniform (probably a spiraling) magnetic field
permeates the jet.

Besides the reference values, we have also explored what range of the parameter 
space is compatible with observations. Fortunately, this range is rather large.
To quantify the last statement, we have found that,
a variation by a factor of two of parameters such as $H$,
$\tau_{\parallel}$, $R_0$, $z_0$ and/or by 10\% of the velocity components of the
electrons or $p$, does not alter our conclusions. 

\section{Results and discussion}

In what follows, we present results for the energy and time-lag spectra, the 
autocorrelation and cross correlation functions, and the power-density spectra.
\begin{figure}
\resizebox{\hsize}{!}{\includegraphics[angle=270]{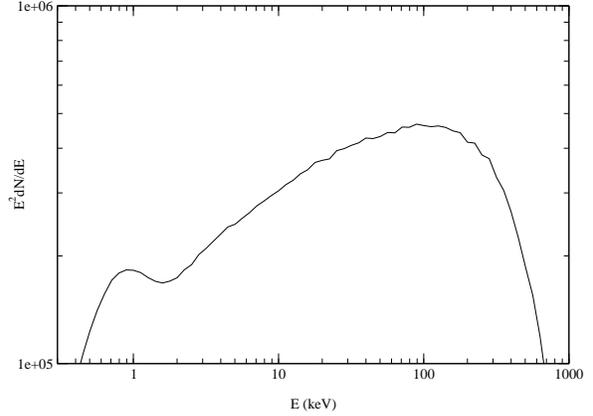}}
\caption[]{Emergent photon-number spectrum multiplied by $E^2$ 
from our jet model for the reference values of the
parameters given in Sect. 3. The soft X-ray excess below 1 keV is essentially
the unscattered soft-photon input. When fitted with a power law with 
an exponential cutoff (see Eq. 6), 
we find $\Gamma=1.7$ and $E_{\rm cut}=280$ keV.}
\label{spectrum}
\end{figure}
\begin{figure}
\resizebox{\hsize}{!}{\includegraphics[angle=270]{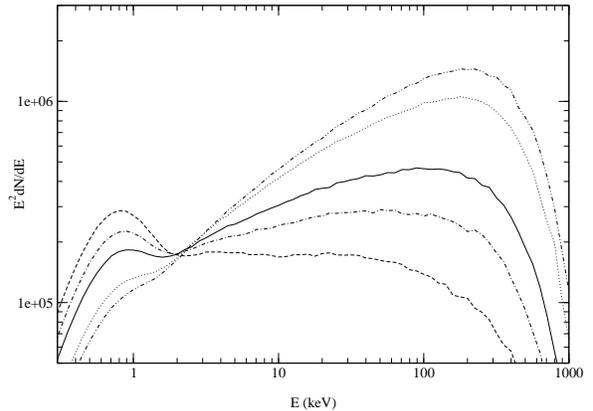}}
\caption[]{Emergent photon-number spectra multiplied by $E^2$
from our jet model for different
optical depths $\tau_{\parallel}$.  The dashed, dash-dotted, solid, dotted
and dash-double-dotted curves correspond to $\tau_{\parallel}=$ 1.5, 2, 2.5, 
3.5, and 4 respectively. The rest of the parameters are kept to their reference values.
The power-law slope decreases with increasing optical depth from
$\Gamma\simeq 2$ (for $\tau_{\parallel}=1.5$) to $\Gamma\simeq 1.45$ (for
$\tau_{\parallel}=4$). The cutoff energy $E_{\rm cut}$ 
varies slowly in the region 260 - 300 keV.  For all the curves, the same
number of input photons $N=4 \times 10^6$ was used.}  
\label{tau}
\end{figure}

\begin{figure}
\resizebox{\hsize}{!}{\includegraphics[angle=270]{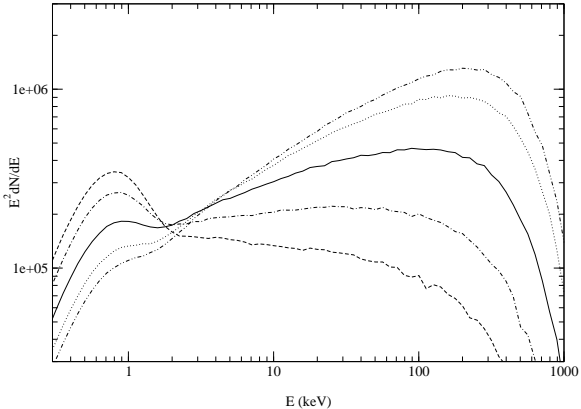}}
\caption[]{Emergent photon-number spectra multiplied by $E^2$
from our jet model for different radii
of the base of the jet. Here the dashed, dash-dotted, solid, dotted,
and dash-double-dotted curves correspond to $R_0/r_g =$  30, 50, 100,
200, and 300 respectively.  The rest of the parameters are kept to their
reference values.  The spectra harden as $R_0$ increases,
because the optical depth to electron scattering
increases along the direction perpendicular to the
axis of the jet. For all the curves, the same number of input photons
$N=4 \times 10^6$ was used.}
\label{R_0}
\end{figure}

\subsection{Energy and time-lag spectra}

The energy spectrum of the emerging photons 
for the reference values of the parameters 
(previous section) is shown in Fig. \ref{spectrum}. The spectrum 
above $\sim 2$ keV can be
well fitted with a power law with an exponential cutoff 
$$ 
dN/dE\propto E^{-\Gamma}\rm e^{-E/E_{\rm cut}},
\eqno(6)
$$
where $\Gamma =1.7$ and $E_{\rm cut}=280$ keV. One can also see the soft 
X-ray spectrum (below $\sim 2$ keV), which is essentially the unscattered
part of the soft-photon input.

Keeping the rest of the parameters to their reference values and varying the
Thomson optical depth along the axis of the jet $\tau_{\parallel}$, we explore
the role of this parameter on the emerging energy spectrum
(Fig. \ref{tau}).  As expected,
higher optical depths result in harder spectra (the photons are scattered more
and gain more energy from the energetic electrons of the jet). The photon 
number spectral index $\Gamma$ decreases from 2 to 1.45, when 
$\tau_{\parallel}$ varies in the range $1.5-4$.

We also made a similar study of the role of the radius $R_0$
of the base of the jet.
By increasing $R_0$, the spectrum hardens and vice-versa (Fig. \ref{R_0}). 
This is so because a
larger $R_0$ translates to a larger Thomson optical depth perpendicular to the
jet axis (we kept the rest of the parameters at their reference values) 
and therefore more photon scatterings.  

 The cutoff energy $E_{\rm cut}$ has a weak dependence on $\tau_{\parallel}$
and $R_0$, but a rather strong dependence on $v_{\perp}$.  This is because
$v_{\perp}$ determines the maximum energy gain of the soft photons.

We then turn our attention to the timing properties that can be extracted
from our model and compare them with the observations. We start with the 
frequency-dependent time lags between two different energy bands. So as to simulate
them, the time of flight of all escaping photons is 
recorded in 16384 time bins
of duration 1/256 s each.  The time of flight is computed by adding up the path
lengths traveled by each photon and dividing the sum by the speed of light.  Then
we consider the light curves of two energy bands,  0.2 - 4 keV and
14 - 45 keV, so as to compare our results with observations directly (e.g. Nowak
et al. 1999). Following Vaughan \& Nowak (1997),
we compute the phase lags and through them the time lags between the two
energy bands as a function of Fourier frequency.

\begin{figure}
\resizebox{\hsize}{!}{\includegraphics[angle=270]{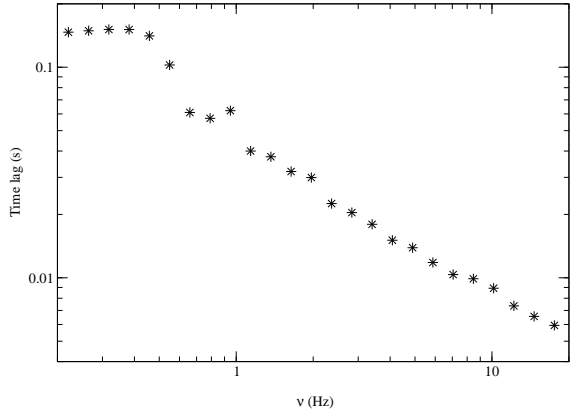}}
\caption[]{Computed time lags as a function of Fourier frequency when the
reference values of the parameters are used.  Positive lags
indicate that hard photons lag soft photons. 
The results are logarithmically rebinned for clarity.
The best-fit
power law (in the 0.2 - 20 Hz frequency range) is $\nu^{-0.8}$.} 
\label{lags}
\end{figure}

Figure~\ref{lags} shows the time lags as a function of Fourier
frequency for the fixed values of the parameters (Sect. 3).
The time-lag spectrum is
approximately represented by a power law $\nu^{-\beta}$,
with index $\beta = 0.8$ in the frequency
range 0.2 - 20 Hz,  in good agreement with the observations of Cyg
X-1 and GX 339--4 (Nowak et al. 1999; Ford et al. 1999).  
The low-frequency ($\sim 0.4$ Hz) break of the time 
lags is determined by the range of directional cosines with respect
to the magnetic field direction in
the jet that we consider (Sect. 3).  A larger range would produce
a break at lower Fourier frequencies and vice versa. 

\begin{figure}
\resizebox{\hsize}{!}{\includegraphics[angle=270]{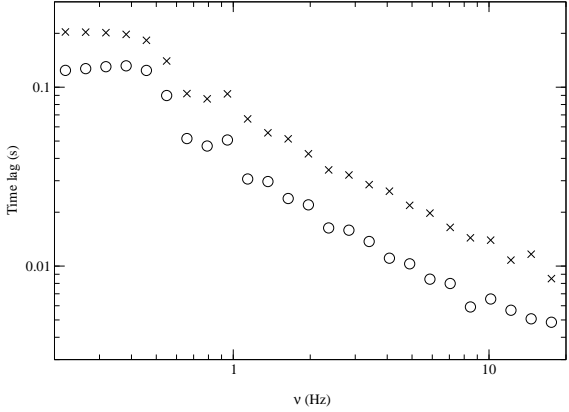}}
\caption[]{Computed time lags as a function of Fourier frequency for different
dimensions of the jet. The crosses correspond to a large jet with $R_0=300r_g$
and $z_0=20r_g$, while the open circles to a smaller jet with $R_0=50r_g$ and
$z_0=2r_g$. The lags of the bigger jet are a factor of 2 larger than these
of the smaller one at the same frequency. In either case, the 
best-fit power law (in the 0.2 - 20 Hz frequency range) is $\nu^{-0.8}$.}
\label{lags2}
\end{figure}

The time lags depend (among other things) on the dimensions of the jet.
The scale of the jet is set by $R_0$ and $z_0$ (see eq. 3).
So as to quantify this dependence, we consider a ``large'' jet ($R_0=300r_g$,
 $z_0=20r_g$) and a scaled down one ($R_0=50 r_g$, $z_0=2r_g$). In Figure 
\ref{lags2} the time lags are presented for these two parameter sets 
and it is obvious that the larger jet results in larger lags by a factor
of 2 for the same frequency. Notice, however, that the frequency dependence
of the lags is very similar. 

Pottschmidt et al. (2000) studied the 
evolution of the lags for different epochs of Cygnus X-1 and noticed
that the lags have the tendency to increase in magnitude during state 
transitions. In the context of our model, this can be understood with 
the association of the state transitions with a more extended jet.

\subsection{Auto- and cross-correlation functions}

Maccarone et al. (2000) computed the autocorrelation functions (ACF) of the 
light curves of Cygnus X-1 in different energy bands and showed that the 
width of the ACF {\it decreases} with photon energy. This imposes strong constraints 
on the different models that attempt to describe both the spectral and the temporal
properties of accreting black-hole binaries.

\begin{figure}
\resizebox{\hsize}{!}{\includegraphics[angle=270]{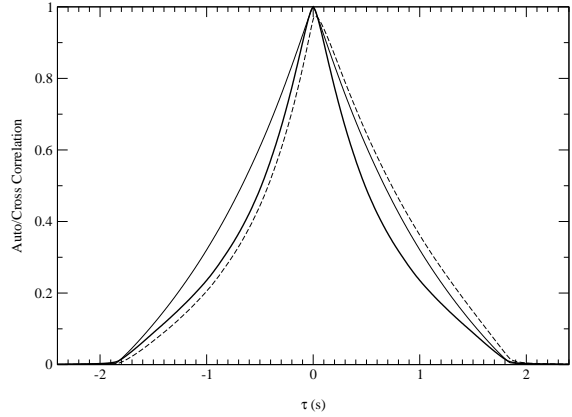}}
\caption[]{The autocorrelation functions of the simulated light curves
in two energy bands is shown. The thin solid curve corresponds to the
soft band (0.2 - 4 keV), while the thick solid curve to the hard band (14 - 45
keV). Both curves are normalized to 1 for lag $\tau =0$. Notice that the 
autocorrelation function becomes narrower in the higher energy band.
We also plot (dashed curve) the cross-correlation function of the two
light curves.}
\label{correl}
\end{figure}

With the simulated light curves of our jet model at hand (for the reference
values of the parameters given in Sect. 3), it is 
straightforward to calculate the ACF at two different energy bands.  
The bands are kept to 0.2 - 4 keV (soft band) and 14 - 45 keV (hard band).
The ACFs are normalized to unity at zero lag in both bands and are shown
in Fig. \ref{correl}.  We find that the 
width of the ACF decreases with photon energy, in qualitative agreement
with observations.  At first sight, this is contrary to intuition.  It can
never happen in a spherical Comptonizing medium.  However, in our model
the high-energy photons are not escaping isotropically, i.e., with a uniform 
distribution in $\cos \theta$.  The more energetic photons come out
preferentially at large values of $\cos \theta$ and with a lightcurve 
which is narrower than that of the less energetic photons.  
This gives rise to a narrowing of the ACF with increasing photon energy.
In a realistic model with a non-uniform magnetic field, this could come 
about as follows:  High-energy photons (escaping with large values of 
$\cos \theta$) come from relatively small regions of the jet and have a 
narrower lightcurve than the softer photons which come from the entire jet.

In Fig. \ref{correl} we also plot the cross-correlation function of
the two light curves. It should be stressed that the exact shape of the 
auto- and cross-correlation functions depends on the properties of the
shots of the soft-photon input.   

\subsection{Power-density spectra} \label{powersp}

Following the method described in Sect. \ref{power}, we calculate the
high-frequency part of the power spectrum in five different photon energy
bands (0.2 - 4 keV, 4 - 6.3 keV, 6.3 - 9 keV, 9 - 14 keV, and 14 - 45 keV).

In the context of our model, the second break in the power spectrum
corresponds to the Keplerian frequency at $\rho=R_0$ in the accretion disk,
which for $R_0 = 100 r_g = 1.5 \times 10^8$ cm is $\nu_{\rm br}\simeq 3$ Hz.
At higher frequencies, the rms soft-input variability is assumed to have a
power-law frequency dependence 
$$
rms_{\rm in} = rms_0 \left( \nu \over {3 {\rm Hz}} \right)^{-1/3},
\eqno(7)
$$ 
so as to match 
the high-frequency slope of the power spectrum for low-energy photons ($E \sim 1$
keV). Note that for  $rms_{\rm in} \propto \nu^{-1/3}$ 
we have $ rms_{\rm in}^2/\rm Hz \propto
\nu^{-5/3}$. Here $rms_0$ is the observed rms at 3 Hz.

\begin{figure}
\resizebox{\hsize}{!}{\includegraphics[angle=270]{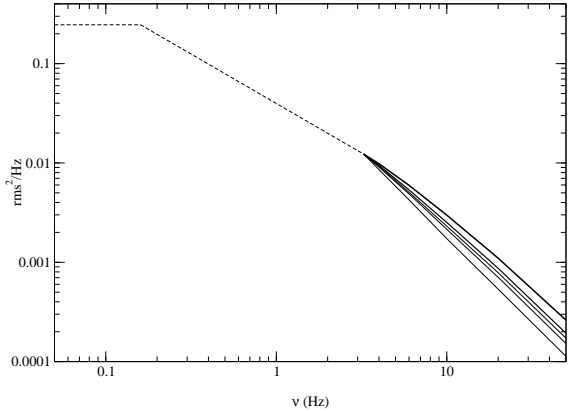}}
\caption[]{Power-density spectrum of Cyg X-1. The dashed line corresponds to
the low-frequency part of the power spectrum, as typically observed. The part of 
the power spectrum above the second break is the result of our simulations and 
shows the spectrum's photon-energy dependence. The five curves correspond 
to different energy bands as follows: $b_1: (0.2 - 4), b_2: (4 - 6.3), 
b_3: (6.3 - 9), b_4: (9 - 14), b_5: (14 - 45)$ keV. 
The steepest slope (thinner curve) corresponds to $b_1$, the 
second in steepness to $b_2$ and so forth.} 
\label{variab}
\end{figure}

We remind the reader that, according to our assumptions (Sect. \ref{power}),
the variability at a given Fourier frequency $\nu$ is identified with the one 
that comes from the part of the accretion disk with $r$ 
constrained by expression (5).  Thus, for a given Fourier frequency $\nu$,
we vary the soft-photon input from the appropriate part of the accretion disk 
by an amount determined by Eq. (7)
and record the resulting variability in the different
energy bands. We find that, the higher the Fourier frequency 
the larger the fractional change in
$rms^2/{\rm Hz}$ between the soft band (0.2 - 4 keV)
and one of the higher bands (Fig. \ref{variab}).
This can be understood as follows: The higher the Fourier frequency the closer to
the core of the jet the source of variability is and thus the soft-input photons
get upscattered in a ``hotter'' medium.  By construction of our model,
the curves of $rms^2/{\rm Hz}$ for the different energy bands meet 
at $\sim \nu_{\rm br}$, which corresponds to $r=R_0$, i.e., the entire 
accretion disk considered in our model.

\begin{figure}
\resizebox{\hsize}{!}{\includegraphics[angle=270]{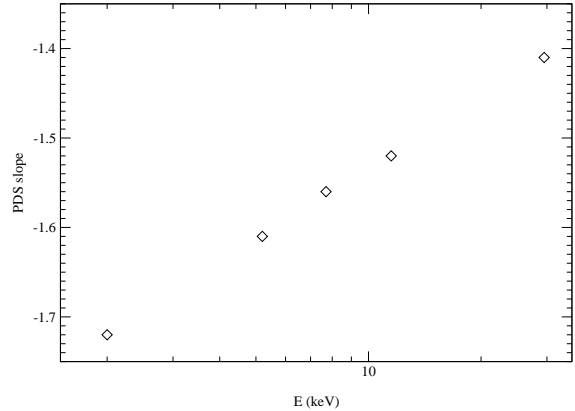}}
\caption[]{Power-law slope of the high-frequency power-density spectra
(Fig. \ref{variab}) versus photon energy.  The abolute value of the slope
for the different energy bands is  
$b_1: 1.72,\quad b_2: 1.61,\quad b_3: 1.56,\quad b_4: 1.52,\quad b_5: 1.41$.}
\label{slope}
\end{figure}

As a consequence of the above, the high-frequency ($\nu > \nu_{\rm br}$)
slope of the power spectrum depends on the energy band. 
A power-law fit to the power spectrum for the soft energy band (0.2 - 4 keV) 
results in an index 1.7, essentially unchanged from the input value 5/3.  
As the band energy increases, the power spectrum flattens and for the
hardest energy band considered (14 - 45 keV) the index becomes $1.4$ 
(Fig. \ref{slope}),
 in agreement with the observations of Cyg X-1 (Nowak et al. 1999). 
This variation of the index by $\sim 0.3$
is a consequence of the negative radial gradient of the perpendicular
component of the electron velocity $v_{\perp}$ and
depends on how much more energetic are the electrons in the core of the jet
in comparison to those at its periphery. Here we have taken
$v_{\perp,0}=0.55 c=2 v_{\perp,{\rm out}}$.   
Had we taken $v_{\perp,0}=v_{\perp,{\rm out}}$ (as in Reig et al. 2003), 
no flattening of the power-density spectra with photon energy would have been 
produced. This may be the case for some black-hole candidates (Lin et al. 1999).    

\section{Conclusion}

The non-thermal X-ray emission in black-hole binaries is usually modeled to 
come from inverse Compton scattering of soft photons in a hot corona (e.g.
Poutanen \& Fabian 1999) that 
lies above and below a thin disk (Shakura \& Sunyaev 1973) or in a 
two-temperature flow (Shapiro et al. 1976). Here we explore an alternative 
picture where the inverse Compton scattering takes place in the jet whose 
presence, whenever the source is in the low/hard state, is well established 
(Fender 2001; Stirling et al. 2001; Gallo et al. 2003).

In a previous work (Reig et al. 2003), it was shown that, for a range of the 
parameters of a very simple jet model, both the energy spectra and the
frequency-dependent time lags of Cygnus X-1 could be reproduced. In this work
we have added a negative gradient along the polar direction $r$ of the electron 
velocity component $v_{\perp}$, resulting in a jet whose core is ``hotter'' than its 
periphery. This jet model can reproduce not only the energy spectrum and the 
time lags
but also the fact that the autocorrelation function is measured to be
narrower (Macarrone et al. 2000) and the high-frequency power spectrum harder 
(Nowak et al. 1999) for the harder photon energy bands. 

In the context of this model the X-ray variability comes from the soft photon
input that is further reprocessed in the jet. The high frequency break in the 
power spectrum (typically seen at $\sim$1 Hz) is identified with the Keplerian
frequency at the outer edge of the base of the jet.

One important challenge to our model is to demonstrate that, with no change 
of the assumptions, one can explain quantitatively the observed radio spectrum
of black-hole sources in the low/hard state 
(Fender 2001) and the correlation between radio and X-ray luminosities
(Gallo et al. 2003).  Preliminary calculations indicate that this is
possible.  A detailed presentation is upcoming (Giannios, in preparation).

Neutron-star systems in the island state exhibit energy spectra and time
variability similar to those of black-hole systems.  We therefore suggest
that our model may be relevant for these neutron-star systems as well.
                      
{\it Acknowledgements.}  One of us (NDK) would like to thank Iosif Papadakis for
useful discussions.  This research has been supported in part by the Program
``Heraklitos'' of the Ministry of Education of Greece.


\begin{thebibliography}{}

\bibitem{} Abramowicz, M. A., Chen, X. M., Kato, S., Lasota, J. P., \& Regev,
O. 1995, ApJ, 438, L37
\bibitem{} Band, D., \& Grindlay, J. E. 1986, ApJ, 311, 595
\bibitem{} Bednarek, W., Kirk, J. G., \& Mastichiadis, A. 1996, A\&AS, 120,
571
\bibitem{} Begelman, M. C., \& Sikora, M. 1987, ApJ, 322, 650
\bibitem{} Belloni, T., Psaltis, D., \& van der Klis, M. 2002, ApJ, 572, 392  
\bibitem{} Cashwell E. D., \& Everett C. J. 1959, {\em A Practical Manual on
the Monte Carlo Method for Random Walk Problems}, Pergamon, Oxford
\bibitem{} Churazov, E., Gilfanov, M., \& Revnivtsev, M. 2001, MNRAS, 321, 759
\bibitem{} Corbel, S. \& Fender, R. P. 2002, ApJ, 573, 35
\bibitem{} Esin, A. A., McClintock, J. E., \& Narayan, R. 1998, ApJ, 500, 523 
\bibitem{} Fender, R. P. 2001, MNRAS, 322, 31
\bibitem{} Fender, R. P.. Pooley, G. G., Brocksopp C., Newell, S. J. 1997, MNRAS,
290, L65 
\bibitem{} Ford, E. C., van der Klis, M., M\'endez, M., van Paradijs, J.,
\& Kaaret, P. 1999, ApJ, 512, L31 
\bibitem{} Galeev, A. A., Rosner, R., \& Vaiana, G. S. 1979, ApJ, 229, 318
\bibitem{} Gallo, E., Fender, R. P., \& Pooley, G. G. 2003, MNRAS, 344, 60 
\bibitem{} Georganopoulos, M., Aharonian, F. A., \& Kirk, J. G. 2002, A\&A,
388, L25
\bibitem{} Haardt, F., \& Maraschi L. 1993, ApJ, 413, 507
\bibitem{} Haardt, F., Maraschi L., \& Ghisellini, G. 1994, ApJ, 432, L95
\bibitem{} Harris, D. E., \& Krawczynski, H. 2002, ApJ, 565, 244
\bibitem{} Hjellming, R.M., \& Han, X. 1995, {\em Radio properties of X-ray binaries.}
In : Lewin, W. H. G., van Paradijs, J., van der Heuvel, E. P. J., (Eds.), X-ray binaries,
Cambridge University Press, Cambridge, 308-330 
\bibitem{} Hua, X., Kazanas, D., \& Titarchuk, L. 1997, ApJ, 482, L57
\bibitem{} Hua, X., Kazanas, D., \& Cui, W. 1999, ApJ, 512, 793
\bibitem{} Ichimaru, S. 1977, ApJ, 214, 840
\bibitem{} Kazanas, D., Hua, X., \& Titarchuk, L., 1997, ApJ, 480, 735
\bibitem{} Kazanas, D., \& Hua, X. 1999, ApJ, 519, 750
\bibitem{} Kylafis, N. D., Giannios, D., \& Psaltis, D. 2004, in {\em X-ray Timing
2003: Rossi and Beyond, eds. P. Kaaret, F.K. Lamb, \& J.H. Swank, in press}, 
\bibitem{} Lehr, D. E., Wagoner, R.V., \& Wilms, J. astro-ph/0004211
\bibitem{} Levinson, A., \& Blandford, R. 1996, ApJ, 456, L29
\bibitem{} Lin, D., Smith, I. A., B$\rm \ddot{o}$ttcher, M., \& Liang, E.P. 
1999, ApJ, 531, 963
\bibitem{} Maccarone, T. J., Coppi, P. S., \& Poutanen, J. 2000, ApJ, 537, L107
\bibitem{} Markoff, S., Falcke, H., \& Fender, R. 2001, A\&A, 372, L25
\bibitem{} Merloni, A., Fabian, A., \& Ross, R. 2000, MNRAS, 313, 193
\bibitem{} Mirabel, I. F., Dhawan, V., Chaty S., et al. 1998, A\&A, 330, L9
\bibitem{} Mitsuda, K., Inoue, H., Koyama, K., et al. 1984, PASJ, 36, 741
\bibitem{} Miyamoto, S., Kitamoto, S., Mitsuda, K., \& Dotani, T. 1988,
Nature, 336, 450
\bibitem{} Miyamoto, S., Kitamoto S., Iga, S., Negoro, H., Terada, K. 1992,
ApJ, 391, L21
\bibitem{} Narayan, R., \& Yi, I. 1994, ApJ, 428, L13
\bibitem{} Nowak, M. A. 2000, MNRAS, 318, 361
\bibitem{} Nowak, M. A., Vaughan, B. A., Wilms, J., Dove, J. B., \& Begelman,
M. C. 1999, ApJ, 510, 874  
\bibitem{} Pottschmidt, K., Wilms, J., Nowak, M. A., et al. 2000, A\&A, 357, L17
\bibitem{} Pottschmidt, K., Wilms, J., Nowak, M. A., et al. 2003, A\&A, 407, 1039
\bibitem{} Poutanen, J. 2001, Adv. Space Res., 28, 267 
\bibitem{} Poutanen, J., \& Fabian, A. C. 1999, MNRAS, 306, L31 
\bibitem{} Pozdnyakov L. A., Sobol I. M., \& Sunyaev R. A. 1983,
Astrophys. \& Space Phys. Rev. 2, 189
\bibitem{} Rees, M. J., Phinney, E. S., Begelman, M. C., \& Blandford, R. P.
1982, Nature, 295, 17 
\bibitem{} Reig, P., Kylafis, N. D., \& Spruit, H. C. 2001, A\&A, 375, 155
\bibitem{} Reig, P., Kylafis, N. D., \& Giannios, D. 2003, A\&A, 403, L15
\bibitem{} Revnivtsev, M., Gilfanov, M., \& Churazov, E. 2000, A\&A, 363, 1013
\bibitem{} Romero, G. E., Kaufman Bernad\'o, M. M., \& Mirabel, F. 2002,
A\&A, 393, L61
\bibitem{} Stern, B. E., Poutanen, J., Svensson, R., Sikora, M., \& Begelman,
M. C. 1995, ApJ, 449, L13
\bibitem{} Stirling, A. M., Spencer, R. E., de la Force C.J., et al. 2001, MNRAS,
327, 1273
\bibitem{} Shakura, N. I., \& Sunyaev, R. A. 1973, A\&A, 24, 337
\bibitem{} Shapiro, S. L., Lightman, A. P., \& Eardley, D. M. 1976, ApJ, 204, 187
\bibitem{} Sunyaev, R. A., \& Titarchuk, L. G., 1980, A\&A, 86, 121
\bibitem{} van der Klis, M. 1995, in {\em The lives of neutron stars},
Kluwer Academic Publishers., Eds. M. A. Alpar, \"U. Kiziloglu, \& J. van Paradijs,
NATO ASI Series C450.
\bibitem{} Vaughan, B. A., \& Nowak, M. A. 1997, ApJ, 474, L43


\end{thebibliography}
\end{document}